\documentclass[onecolumn]{article}
\pdfoutput=1
\usepackage{graphicx}
\usepackage{amsmath}
\usepackage{amssymb}
\usepackage{bigstrut}
\usepackage{tabularx,ragged2e,booktabs,caption,authblk}
\newcolumntype{C}[1]{>{\Centering}m{#1}}

\everymath{\displaystyle}
\usepackage{fullpage}
\usepackage{cite}
\usepackage{float}

\title{Asynchrony promotes polarized collective motion in attraction based models}

\author[1,2,3]{Daniel Str{\"{o}}mbom}
\author[2]{Tasnia Hassan} 
\author[2]{W Hunter Greis}
\author[4]{Alice Antia}

\affil[1]{Department of Mathematics, Uppsala University, Uppsala, Sweden}
\affil[2]{Department of Biology, Lafayette College, Easton, PA, USA}
\affil[3]{Department of Biosciences, Swansea University, Swansea, UK}
\affil[4]{Department of Mathematics and Statistics, Carleton College, MN, USA}

\date{}

\begin{document}

\maketitle

\begin{abstract}
\noindent Animal groups frequently move in a highly organized manner, as represented by flocks of birds and schools of fish. Despite being an everyday occurrence, we do not yet fully understand how this works. What type of social interactions between animals gives rise to the overall flock structure and behavior we observe? This question is often investigated using self-propelled particle models where particles represent the individual animals. These models differ in the social interactions used, individual particle properties, and various technical assumptions. One particular technical assumption relates to whether all particles update their headings and positions at exactly the same time (synchronous update) or not (asynchronous update). Here we investigate the causal effects of this assumption in a specific model and find that it has a dramatic impact. In particular, polarized groups do not form when synchronous update is used, but are always produced with asynchronous updates. We also show that full asynchrony is not required for polarized groups to form and quantify time to polarized group formation. Since many important models in the literature have been implemented with synchronous update only, we speculate that our understanding of these models, or rather the social interactions on which they are based, may be incomplete. Perhaps a range of previously unobserved dynamic phenomena will emerge if other potentially more realistic update schemes are chosen. 
\end{abstract}

\section*{Introduction}
Moving animal groups, such as schools of fish and flocks of birds, often move in a highly coordinated fashion despite the fact that each member of the group only experiences its immediate surroundings and often no leader can be identified. How does that work?

This question is typically investigated using self-propelled particle (SPP) models. In a typical SPP model a number of particles move in the plane, or space, and update their headings at each time step according to a specified local interaction rule operating on the position, and/or the heading, of nearby particles. An example of a common local interaction rule is one where particles are repelled from nearby particles (repulsion), take the average heading of particles at intermediate distances (orientation), and are attracted to particles which are further away (attraction) \cite{Reynolds1987,Couzin2002}. A large number of models implementing various subsets of attraction, repulsion and orientation have been proposed and analyzed in recent years \cite{Vicsek2012,Strombom2015,Barberis2016}.

SPP models have proven successful in explaining how collective motion may emerge from repeated local interactions between individuals in general settings and in specific experimental and real world situations \cite{Vicsek2012}. However, the use of SPP models has also attracted some criticism, both as models of real world phenomena and in relation to how they are constructed \cite{Hemelrijk2008,Lopez2012,Fine2013}. Over the past decade, models have been adapted in various ways to resolve some of the issues. For example, models have included non-constant speeds \cite{Hemelrijk2008}, more realistic neighbor detection \cite{Ginelli2010,Lemasson2013,Kunz2012}, asynchronous updates \cite{Bode2010a,Strombom2011}, more realistic visual system \cite{Lemasson2009}, leaders and shepherds \cite{Couzin2005,Strombom2014,Ferdinandy2017}, explicit environmental and social coupling \cite{Guttal2010}, and much more. 

Although other specific concerns are outlined in \cite{Hemelrijk2008,Lopez2012,Fine2013} most that have been addressed focus on improving the models by making some aspect of the individual particles more realistic. What about more low-level assumptions and choices? For example, regardless of how sophisticated the individuals and the social interactions between them are, all SPP models must include instructions for how to update particle headings and positions. One option is to update all particle positions and headings at exactly the same time (synchronous update), or use some type of asynchronous update scheme where particles may update their headings and positions at different times. This issue has been thoroughly investigated in related fields and shown to be important. For example in robotics \cite{Beni2004,Flocchini2008,Liu2003,Samilouglu2006} and cellular automata \cite{Ingerson1984,Bersini1994,Schonfisch1999,Fates2013,Sethi2016}. In particular, direct comparison of asychronous and synchronous versions of particular cellular automata show that asynchronous update tend to increase the stability of the automaton, see \cite{Fates2013} for an overview. 

Direct comparisons of this type are largely absent from the SPP model literature, and in most models it is assumed that all particles calculate and update their headings synchronously. This assumption has been questioned by several authors and some have chosen to implement asynchronous update schemes \cite{Bode2010a,Bode2010b,Strombom2011}. These studies have revealed that implementing an asynchronous update scheme allows for some previously elusive empirical observations to be reproduced by SPP models. In particular, speed distributions in fish schools \cite{Bode2010a} and interactions of a topological nature consistent with those observed in starling flocks \cite{Bode2010b}. However, despite these particular empirically motivated findings, systematic direct comparisons of various update schemes in standard SPP models have not been conducted, and it is still largely unknown what effects the choice of update scheme may have on models of this type. From a mathematical/computational point of view this choice may well have a dramatic effect. Potentially in a way similar to other well documented choices made in model construction. For example, discrete-continuous and spatial-nonspatial \cite{Durrett1994}. On the other hand, many seemingly different SPP models produce very similar behavior at the collective level \cite{Vicsek2008,Vicsek2012}. This suggests that perhaps these models are more robust to changes in underlying assumptions and detailed implementation choices than more traditional dynamical systems.

Here we compare the effect of implementing the synchronous and a particular asynchronous update scheme in the simplest SPP model known to produce the three standard groups (polarized groups, mills, and swarms), the local attraction model (LAM) \cite{Strombom2011}. If this choice has an effect on this model it is likely to have an effect on more sophisticated models, and as much of our current understanding of collective motion in moving animal groups is based on SPP models, this would be a valuable insight.

\section*{Model and Simulations}
In its simplest form, the LAM is a self-propelled particle model in which $N$ particles move at constant speed in two dimensions and interact via local attraction only. See Fig 1A. In every time step, each particle calculates the position of the local center of mass (LCM) of all particles within a distance of $R$ from it. The particle’s new heading ($\bar{D}_{t+1}$) is a linear combination of the normalized direction toward the local center of mass ($\hat{C}_{t}$) and its previous direction ($\hat{D}_{t}$). The parameter $c$ specifies the relative strength of attraction to the local center of mass, and the parameter $d$ the relative tendency to proceed in the previous direction. The particle will then move a distance of $\delta$ in the direction specified by $\bar{D}_{t+1}$. From \cite{Strombom2011} we know that different groups will form depending on if $c$ is less than, approximately equal to, or larger than $d$. More specifically, if $c<<d$ polarized groups form (Fig 1Ba). If $c\approx d$ mills will form (Fig1Bb). If $c>>d$ swarms will form (Fig 1Bc). At least it is when an asynchronous update scheme is used.

\begin{figure}[tbhp]
\centering
\includegraphics[width=.75\linewidth]{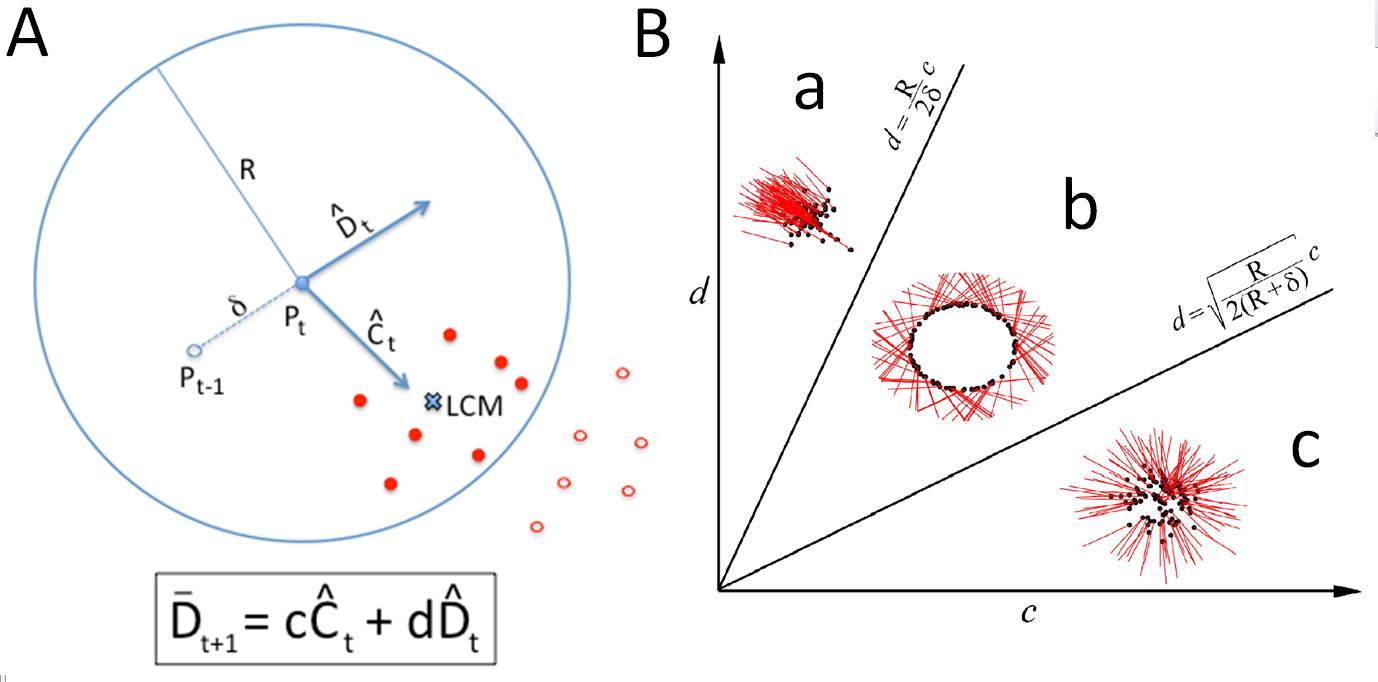}
\caption{(A) Illustration of how each particle calculates its new heading on each time step. (B) The groups present in the simplest version of the LAM model (Reproduced from \cite{Strombom2011}). (Ba) Polarized group, (Bb) Mill, and (Bc) Swarm.}
\label{fig:Fig1a}
\end{figure}

\subsection*{Synchronous and asynchronous update schemes}
The synchronous update scheme is the standard updating scheme where all particles calculate their new headings and update their positions at exactly the same time. The asynchronous update scheme chosen here is one in which all the particles update their headings and positions sequentially in each time step, and the order in which they do so is randomized from one time step to the next. This particular choice is motivated in the discussion.

\subsection*{Simulations} 
First, we ran 100 simulations for each $c$ from 0.04 to 2 in increments of 0.02 in both the synchronous case and in the asynchronous case. Keeping all other parameters fixed at $d=1$, $N=50$, $R=4$, $\delta=0.5$. Periodic boundary conditions were used and each particle was assigned a random position and heading at the start of each simulation. We measured the polarization ($\alpha$) and scaled size ($\sigma$) of the resulting group and the number of time steps it took for the group to form $\tau$. See the Materials and Methods section for more details. The polarization measures the degree to which the $N$ particles are moving in the same direction \cite{Vicsek1995}. Polarized groups have $\alpha \approx 1$, mills have $\alpha \approx 0$. Swarms and random configurations have intermediate $\alpha$ values . Scaled area provides a measure of how much of the available space the group occupies \cite{Strombom2011}. If no group formed $\sigma$ is large ($\approx 1$), polarized groups have small $\sigma$, mills have a value that decreases with $c$, and swarms have very small $\sigma$. Combining these two measures allows us to distinguish between the three groups in Fig 1B and the case when no group has formed. 
To investigate the formation of polarized groups in more detail we ran a set of simulations containing a mix of $N_{a}$ asynchronously updating particles and $N-N_{a}$ synchronously updating particles with $c=0.1$. More specifically, we ran 100 simulations for each $(N,N_{a})$-pair with $N$ from 2 to 102 in increments of 10, and $N_{a}$ from 0 to $N$ in increments of 2.


\section*{Results}
Depending on the choice of asynchronous or synchronous update alone, the tendency of the model to produce polarized groups (Fig 1Ba) is very different. In Fig 2, we see that for $c$ less than 0.2 there is a dramatic difference between the asynchronous update and synchronous update. In the former, we see the signature of polarized groups (high $\alpha$, low $\sigma$) and in the latter, no group (low $\alpha$, high $\sigma$). Fig 3 shows the time to polarized group formation for $c$ from 0.04 to 0.18 over 100 simulations in the asynchronous case. We see that the time to formation decreases with $c$ and tend to be less than 10000 time steps. Corresponding simulations in the synchronous case with an upper time limit of $10^8$ time steps did not produce a single polarized group. Continuing on Fig 2 for $c$ from just above 0.2 to 1 the asynchronous and synchronous update produce very similar results, this is the mill regime and it appears largely unaffected by the update scheme choice. For $c$ larger than 1 the asynchronous and synchronous case produce qualitatively different results. In the asynchronous case mills quickly degenerate as $c$ increases above 1 and from then on more or less mobile swarms are produced. In the synchronous case mills are being produced for $c$ larger than 1, only becoming truly degenerate and swarm-like very close to $c=2$. 

\begin{figure}[tbhp]
\centering
\includegraphics[width=0.75\linewidth]{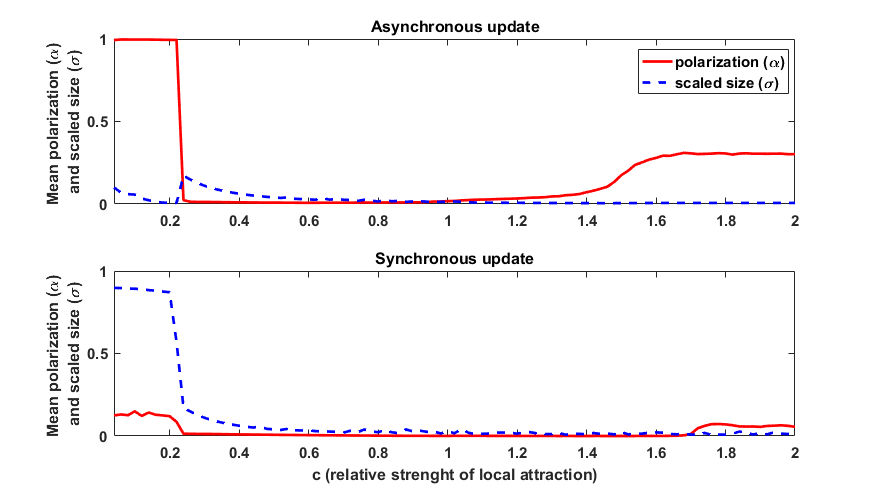}
\caption{Mean polarization and scaled size over 100 simulations for each $c\in[0.04,2]$ with asynchronous update (top) and synchronous update (bottom).}
\label{fig:Fig2}
\end{figure}
\begin{figure}[tbhp]
\centering
\includegraphics[width=0.75\linewidth]{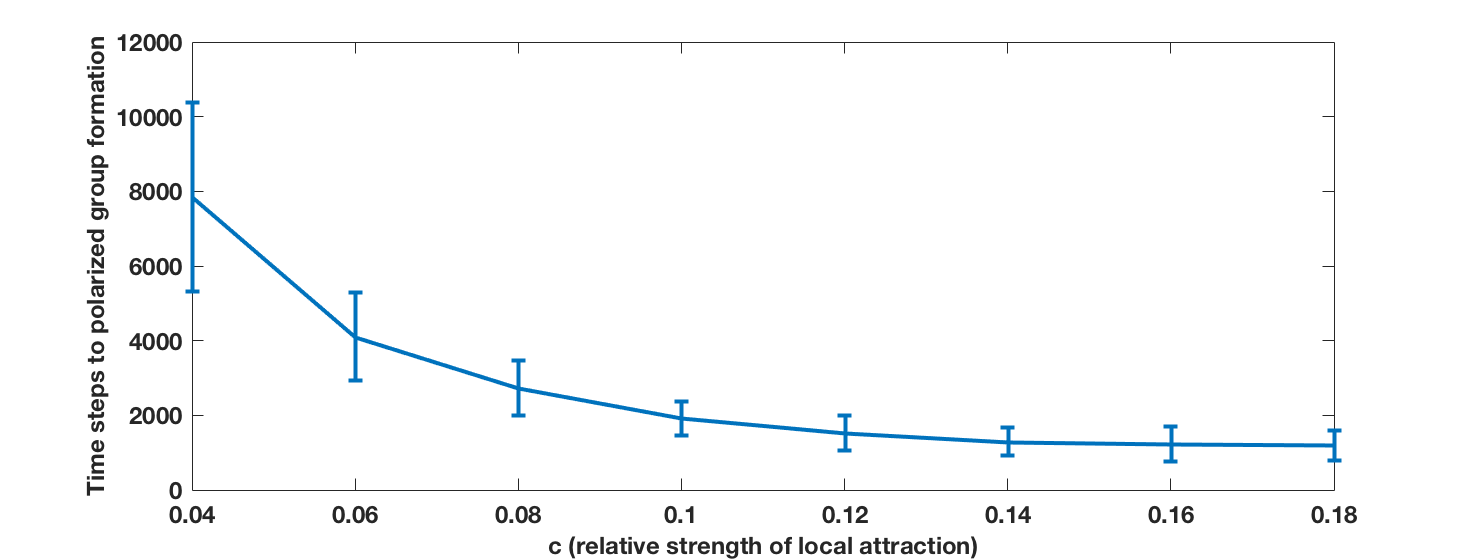}
\caption{Time to polarized group formation over 100 simulations for each $c\in[0.04,0.18]$ with asynchronous update. The line represents the mean and the bars the standard deviations.}
\label{fig:Fig3}
\end{figure}

We also established that the production of polarized groups does not require complete asynchrony in updates. Rather a certain proportion of the particles must update asynchronously for polarized groups to form. In fact, as Fig 4 shows, polarized groups realiably form even when a majority of the particles are updating synchronously for groups larger than or equal to 12. At $N=102$, if the proportion of asynchronously updating particles is larger than approximately 1/5, then polarized groups form. We also note that the proportion of asynchronously updating particles required for polarized group formation decreases with $N$.

\begin{figure}[tbhp]
\centering
\includegraphics[width=0.75\linewidth]{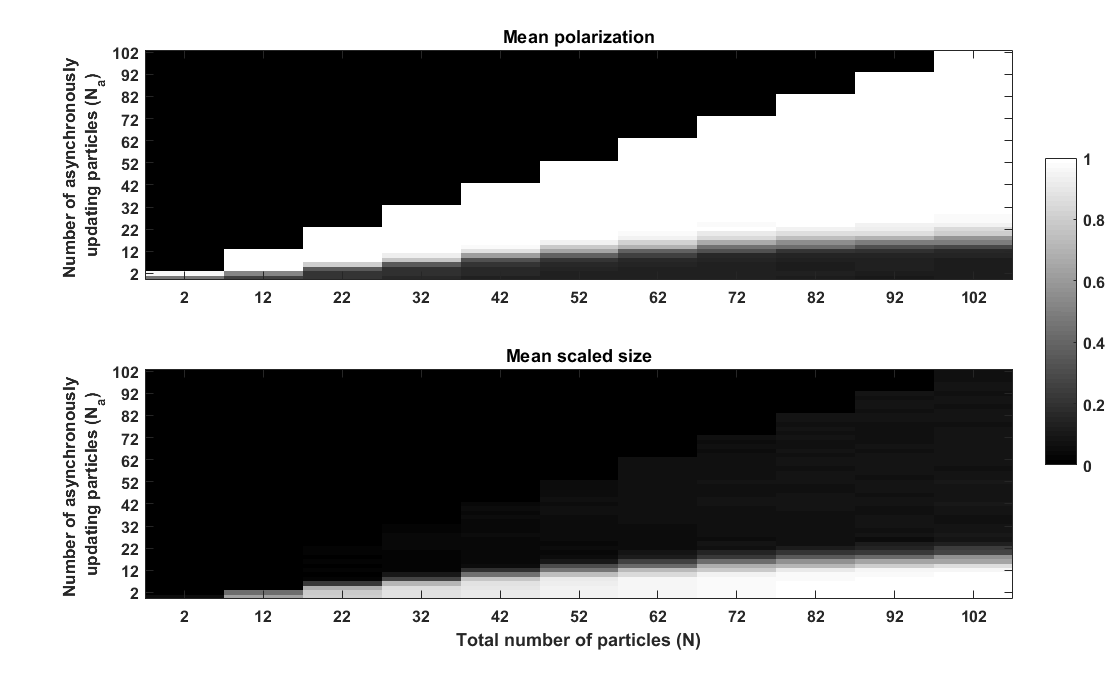}
\caption{Mean polarization and scaled size over 100 simulations for each $(N,N_a)$-pair.}
\label{fig:Fig4}
\end{figure}

\section*{Discussion}
We have shown that the choice between asynchronous and synchronous update has a dramatic effect on the simplest version of the local attraction model. In particular, the formation of polarized groups is inhibited in the synchronous case (Fig 2). Considering that many influential models of collective motion exclusively use synchronous update this suggests that further analysis of these models may reveal previously unseen groups and collective phenomena. Adding to the asynchrony induced speed distribution and topological-like interactions described in \cite{Bode2010a,Bode2010b}. How a model will behave under the local interaction rules alone when other potentially more realistic update schemes are used is largely unknown. Perhaps properties that are rare, or non-existent, in groups generated by standard SPP models, for example the multistability and transition behavior reported in \cite{Tunstrom2013}, are sensitive to choice of updating scheme. In particular, as it appears that the synchronous update may contribute to calm the system down, as exemplified by the swarm phase in our study. In the asynchronous case swarms are more mobile, as indicated by high polarization for large $c$ in Fig 2 (top). In contrast in the synchronous case the swarms are almost stationary, exhibiting very low polarization for large $c$ in Fig 2 (bottom). 

We also show that polarized groups reliably form in simulations with a mix of asynchronously and synchronously updating particles, and the proportion of asynchronously updating particles required decreases with the total number of particles (Fig 4). Perhaps this may have some relevance to modeling leadership in moving animal groups. At least it does not seem unreasonable that leaders, or informed individuals, would update more asynchronously as they know where to go or what to do, whereas followers may decide to update more synchronously, especially in a quorum decision making type situation.

Despite being criticized, the synchronous update still seems to be the default choice in SPP model construction. We speculate that there are many reasons for this. For example, most well-known models were originally presented in that way and it is more straight forward to obtain a continuum approximation of the model and thus make other analysis tools available. In addition, if one decides to use asynchronous updates, which particular scheme should one choose? The asynchronous update scheme used here, and in part in \cite{Strombom2011}, was chosen mainly because it has the same update rate at the time step level as the synchronous update against which it was to be compared, the randomization of update sequence between time steps prevents artefacts arising from a strict persistent update order, and it was easy to implement in a way that allowed for a mix of asynchronously and synchronously updating particles. 

It should be emphasized that we are not claiming that the asynchronous update scheme used here is more realistic than the synchronous one against which it was compared. We do claim that this choice in itself may be critically important in the study of collective motion via SPP models. Perhaps, in some situations, it may be as important as the form of the social interaction rule itself, as in our example presented here. Therefore, we suggest that in future theoretical studies of minimal SPP models a variety of different update schemes be explored, presented and compared. In modeling specific experiments, or observations, we suggest using pilot data to estimate the update distribution and use that to inform the update choice selection. It would be very unfortunate if a carefully designed model that has the capacity to reproduce key properties observed in a specific experiment is abandoned, or made more complicated by adding more social interactions or constraints, because an underlying assumption like this one was overlooked. 

\section*{Materials and Methods}

\subsection*{Calculating polarization and scaled size}
The polarization (or alignment) $\alpha$ of a group of $N$ particles at time $t$ is given by $$\alpha=\frac{1}{N}\left|\sum_{i=1}^{N}\hat{D}_{i,t}\right|,$$ where $\hat{D}_{i,t}$ is the normalized heading of particle $i$ at time $t$. It measures the degree to which the particles in the group are moving in the same direction.\\
\\
The scaled size $\sigma$ of a group of $N$ particles distributed on a square of side length $L$ at time $t$ is given by $$\sigma=\frac{(max(P_x)-min(P_x))(max(P_y)-min(P_y))}{L^{2}},$$ where $P_x$ is the set of $x$-coordinates, and $P_y$ is the set of $y$-coordinates, of the particles on the square at time $t$. It measures the area of the smallest square enclosing all particles relative to the area of the whole square, and thus provide an estimate of how much of the available space the group of particles occupy at a certain time \cite{Strombom2011}. Note that when periodic boundary conditions are used the above formula alone is not sufficient to determine the area of the smallest square enclosing all the particles. Because the smallest square may extend over one, or both, boundaries. We deal with this is by calculating the area over all possible boundary crossing cases and then take the minimum. 

\subsection*{Termination conditions for simulations}
A simulation will terminate when the maximum time $T$ is reached, or when a specific group type has been identified. In this article $T=15000$ timesteps, except when investigating polarized group formation in the synchronous case in where it is set to $T=10^8$. To save time a simulation may terminate before the maximum time is reached if a certain group has formed early. This is determined by continuously measuring $\alpha$ and $\sigma$ throughout the simulation and comparing them to values associated with known groups. More specifically, if over 50 consecutive timesteps $\sigma<0.01$ a cohesive group has formed, it can be either a mill, swarm or polarized group and its $\alpha$ value will inform us about which it is. In particular, if $\alpha>0.995$ a polarized group has formed. This allows us to detect most quickly forming groups, except larger mills, so the following additional termination criteria was introduced to detect these. If over 50 consecutive timesteps $\alpha<0.02$ and $0.01<\sigma<0.25$ a larger mill has been detected. 
Once a simulation has terminated, either by reaching the maximum time or terminating early as previously described, the following measurements are returned. The mean of $\alpha$ and $\sigma$ over the last 50 steps of the simulation, and the number of time steps until the simulation terminated $\tau$.

\section*{Acknowledgments}
This work was supported by a grant from HHMI to Lafayette College under the Precollege and Undergraduate Science Education Program, and a grant from the Swedish Research Council to D.S. (2015-06335).

\bibliographystyle{vancouver}
\bibliography{bibsynch}

\end{document}